\documentclass[
	aps,
	amsmath,amssymb,
	reprint,
	superscriptaddress,
	longbibliography
   ]{revtex4-1}
\newcommand{\expect}[1]{
	\ensuremath{\langle{#1}\rangle}
}
\newcommand{\Tr}{\mathrm{Tr}}

\makeatletter
\newsavebox{\@brx}
\newcommand{\llangle}[1][]{\savebox{\@brx}{\(\m@th{#1\langle}\)}
  \mathopen{\copy\@brx\kern-0.5\wd\@brx\usebox{\@brx}}}
\newcommand{\rrangle}[1][]{\savebox{\@brx}{\(\m@th{#1\rangle}\)}
  \mathclose{\copy\@brx\kern-0.5\wd\@brx\usebox{\@brx}}}
\makeatother

\newcommand{\kket}[1]{
	\ensuremath{|{#1}\rrangle}
}
\newcommand{\bbra}[1]{
	\ensuremath{\llangle{#1}|}
}
\newcommand{\bbraket}[1]{
	\ensuremath{\llangle{#1}\rrangle}
}
\newcommand{\red}[1]{ #1 }

\usepackage{graphicx}
\usepackage{color}
\usepackage{bm}
\usepackage{braket}
\usepackage{ulem}
\newcommand{\removed}[1]{
}
\newtheorem{theorem}{Theorem}
\newtheorem{lemma}[theorem]{Lemma}
\begin{document}

\title[]{Constructing a virtual two-qubit gate by sampling single-qubit operations}

\author{Kosuke Mitarai}
\email{mitarai@qc.ee.es.osaka-u.ac.jp}
\affiliation{Graduate School of Engineering Science, Osaka University, 1-3 Machikaneyama, Toyonaka, Osaka 560-8531, Japan.}
\affiliation{Center for Quantum Information and Quantum Biology, Institute for Open and Transdisciplinary Research Initiatives, Osaka University, Osaka 560-8531, Japan.}
\affiliation{JST, PRESTO, 4-1-8 Honcho, Kawaguchi, Saitama 332-0012, Japan}

\author{Keisuke Fujii}
\email{fujii@qc.ee.es.osaka-u.ac.jp}
\affiliation{Graduate School of Engineering Science, Osaka University, 1-3 Machikaneyama, Toyonaka, Osaka 560-8531, Japan.}
\affiliation{Center for Quantum Information and Quantum Biology, Institute for Open and Transdisciplinary Research Initiatives, Osaka University, Osaka 560-8531, Japan.}
\affiliation{Center for Emergent Matter Science, RIKEN, Wako Saitama 351-0198, Japan}

\date{\today}

\begin{abstract}
We show a certain kind of non-local operations can be simulated by sampling a set of local operations with a quasi-probability distribution when the task of a quantum circuit is to evaluate an expectation value of observables.
Utilizing the result, we describe a strategy to decompose a two-qubit gate to a sequence of single-qubit operations.
Required operations are projective measurement of a qubit in Pauli basis, and $\pi/2$ rotation around x, y, and z axes.
The required number of sampling to get an expectation value of a target observable within an error of $\epsilon$ is roughly $O(9^k/\epsilon^2)$, where $k$ is the number of ``cuts'' performed.
The proposed technique enables to perform ``virtual'' gates between a distant pair of qubits, where there is no direct interaction and thus a number of swap gates are inevitable otherwise.
It can also be utilized to improve the simulation of a large quantum computer with a small-sized quantum device, which is an idea put forward by [Peng, \textit{et al.}, arXiv:1904.00102].
This work can enhance the connectivity of qubits on near-term, noisy quantum computers.
\end{abstract}

\pacs{Valid PACS appear here}
\maketitle

\section{Introduction}
Quantum computers have attracted much attention recently, mainly due to the rapid development of actual hardware \cite{Barends2014, Bernien2017, Wright2019}.
The quantum computer that is to appear shortly is called noisy intermediate scale quantum devices, or in short, NISQ devices \cite{Preskill2018}.
We expect NISQ devices to have $\sim$100 of qubits with non-negligible noise in the near future.
Such devices are believed to be not simulatable by classical computers when the control precision of the qubits is sufficiently high \cite{Harrow2017, Boixo2018, Neill2018, Bravyi2018}.
In this sense, NISQ devices have computational power that exceeds classical computers.
Many researchers are actively developing ways to exploit their power for practical applications \cite{Peruzzo2014, Kandala2017, Nam2019, Farhi2014, Otterbach2017, Mitarai2018, Havlicek2019}.
However, we still suffer from the limited number of qubits available on actual devices and the limited depth of circuits that can be run while maintaining the resultant quantum state meaningful.

If techniques to decompose a quantum circuit to a smaller one are developed, they can extend the applicability of such devices.
Smaller quantum circuits may refer to ones with the smaller number of qubits or gates.
Peng \textit{et al.} recently proposed a clustering approach based on a tensor network representation of a quantum circuit~\cite{Peng2019}, which greatly progressed the technical development.
They showed that we can ``cut'' an identity gate, by sampling measure-and-prepare channels on a qubit according to a certain quasi-probability distribution.
In Ref.~\cite{Mitarai2019}, we proposed methods to construct quantum circuits equivalent to the Hadamard test, which successfully reduces the depth of certain quantum circuits.
These techniques share a same idea in that they reconstruct a result of a coherent quantum operation from certain incoherent operations by combining the results obtained from them.

An approach which has the same flavor as the above have been utilized in the context of memory-efficient classical simulation of quantum circuits.
Since the direct simulation of a quantum circuit with over 50 qubits breaks down due to the need of storing $2^{50}$ complex numbers in memory, the classical simulator must decompose the given quantum circuit to smaller ones, especially in the number of qubits.
Refs. \cite{Chen2018, Pednault2017} have provided one way for such decomposition, which ``cuts'' controlled-Z gates by separately simulating two cases where the control qubit is $\ket{0}$ or $\ket{1}$ and then combining them, and they performed classical simulation of over 50-qubit quantum circuits.
A similar technique has been utilized by Bravyi \textit{et al.} in Ref. \cite{Bravyi2016} to remove a relatively small number of qubits from a large quantum circuit by replacing the qubits with a classical simulator.
Their approach can be viewed as ``space-like'' cut rather than the ``time-like'' cut proposed by Peng \textit{et al} \cite{Peng2019}.
However, their techniques are intended to run on a classical computer and cannot be utilized for simulating a large quantum circuit with a small quantum computer.

In this work, we present a technique to perform ``space-like'' cut on a quantum computer.
More specifically, we present a way to decompose a controlled gate into a sequence of single-qubit operations which consists of projective measurements of Pauli $X$, $Y$, and $Z$ operators, and single-qubit rotations around x, y, and z-axes.
We note that our method does not generate any entanglement between the qubits as it is impossible to do so with such single-qubit operations.
Our method only ``simulates'' effects of entanglement using classical post-processing and sampling.
More concretely, although entangling gates cannot be performed with local operations and classical communications in single-shot experiments as widely known \cite{PhysRevA.52.3457}, we show that it is possible to perform a computational task of evaluating expectation values of the output of entangling circuits by sampling certain sets of gates and applying classical post-processing.
The overhead required for our proposed technique, which scales exponentially to the number of decomposition performed, gives a characterization of the entangling gates from a computational viewpoint, which is different from the existing theories of entanglement quantification in e.g. \cite{Vidal_2000}.

The method proposed here can be considered as a generalization of our previous work \cite{Mitarai2019} and a variant of the quantum circuit decomposition presented in Ref.~\cite{Peng2019}.
It can also be viewed as a fully quantum version of the technique utilized in efficient classical simulation schemes \cite{Chen2018, Pednault2017, Bravyi2016}. 
In some cases, our method provides a better scaling against Ref. \cite{Peng2019} when simulating a large quantum circuit with smaller ones.
The proposed technique is also useful when we want to apply two-qubit gates between a distant pair of qubits, which otherwise would require many swap operations to perform.
This work extends the applicability of NISQ devices whose circuit depth and connectivity are limited.

\section{Gate decomposition}
\subsection{Tensor network representation of quantum circuits}
Quantum computation is completely specified with a quantum circuit, $U$, an initial state with its density matrix representation, $\rho$, and an observable, $O$, measured at the output.
Given $U$, $\rho$, and $O$, Any quantum computation can be represented by a tensor network \cite{Shi2006,Markov2008,Vidal2003}.
We define the tensor representation of $U$, $\rho$, and $O$ in the following manner.

Suppose that our quantum computer has $n$ qubits.
We define a complete set of basis in the space of $2 \times 2$ complex matrix and its dual as $\{\kket{e_i}\}_{i=1}^{4}$ and $\{\bbra{e_i}\}_{i=1}^{4}$ respectively, and assume orthonormality under the trace inner product; $\bbraket{e_i|e_j}=\delta_{ij}$.
We use the trace inner product, that is, for matrices $A$ and $B$, $\bbraket{A|B} = \Tr(A^\dagger B)$.
A density matrix $\rho$ can be decomposed into the sum of $\kket{e_{j_1}}\otimes \kket{e_{j_2}} \otimes \cdots \otimes \kket{e_{j_n}} = \kket{e_{j_1}e_{j_2}\cdots e_{j_n}}$ as 
\begin{align}
    \kket{\rho} = \sum_{j_1,\cdots, j_n} \rho_{\bm{j}} \kket{e_{j_1}e_{j_2}\cdots e_{j_n}},
\end{align}
where $\bm{j} = (j_1,j_2,\cdots,j_n)$. 
We refer to the elements $\rho_{\bm{j}} = \bbraket{e_{j_1}e_{j_2}\cdots e_{j_n}|\rho}$ as the tensor representation of $\rho$.
An observable $O$ can also be decomposed into the same form.
Note that we can naturally assume tensor representations of observables and density matrices consist of real numbers because they are always Hermitian and we can choose the basis $\{\kket{e_i}\}_{i=1}^{4}$ as Hermitian, e.g., we can use the Pauli matrices $\{I, X, Y, Z\}$ as the basis.
Therefore, we assume $\rho_i$ and $O_i$ are real henceforth.
The quantum circuit, $U$, transforms $\rho$ into $U\rho U^\dagger$.
We define a corresponding superoperator $\mathcal{S}(U)$ whose action is defined by $\mathcal{S}(U) \rho  = U\rho U^\dagger$.
Superoperator can be decomposed as,
\begin{align}
    \mathcal{S}(U) = \sum_{j_1,\cdots, j_n}\sum_{k_1,\cdots, k_n} \mathcal{S}(U)_{\bm{j}, \bm{k}} \kket{e_{j_1}\cdots e_{j_n}}\bbra{e_{k_1}\cdots e_{k_n}}.
\end{align}
Note that this decomposition is not limited to superoperators of unitary matrices, but also is applicable for any linear operator that acts on a density matrix.
We call $\mathcal{S}(U)_{\bm{j}, \bm{k}} = \bbra{e_{j_1}\cdots e_{j_n}}\mathcal{S}(U)\kket{e_{k_1}\cdots e_{k_n}}$ tensor representation of $\mathcal{S}(U)$. 
When we use the Pauli operators as basis set, $\mathcal{S}(U)_{\bm{j}, \bm{k}}$ is refered as Pauli transfer matrix.

Quantum computation ends with measuring the observable $O$.
This output can be written down as,
\begin{align}
    \bbra{O}\mathcal{S}(U)\kket{\rho} &= \Tr(O U\rho U^\dagger) \\
    &= \sum_{j_1,\cdots, j_n}\sum_{k_1,\cdots, k_n} O_{\bm{j}}\mathcal{S}(U)_{\bm{j}, \bm{k}} \rho_{\bm{k}},
\end{align}
In many cases, $U$ is a product of elementary gates $\{U_i\}_{i=1}^L$, that is, $U=U_L\cdots U_1$.
The tensor representation of the overall gate, $\mathcal{S}(U)$, is also a product of $\mathcal{S}(U_i)$; $\mathcal{S}(U) = \mathcal{S}(U_L)\cdots\mathcal{S}(U_1)$.
An important note is that as long as the tensor representation of each element is unchanged, the result of the overall computation is also unchanged.
If $\mathcal{S}(U)$ can be represented by a sum of some simple operations as $\mathcal{S}(U) = \sum_i c_i \mathcal{S}(V_i)$ with coefficients $\{c_i\}$, the expectation value of an observable $O$ can be computed with the following equality,
\begin{equation}\label{eq:superop_sum_decomposition}
    \bbra{O}\mathcal{S}(U)\kket{\rho} = \sum_i c_i \bbra{O}\mathcal{S}(V_i)\kket{\rho}.
\end{equation}
Note that $c_i$ can, in general, depend on the state $\kket{\rho}$.
We use this scheme to perform the ``decomposition'' of a circuit in this work.

It is noteworthy that as we perform decompositions of a superoperator rather than an operator such as $U$ itself, the method becomes friendly for a realistic quantum device.
A direct decomposition of $U$ into some simple operators $\{V_i\}$, i.e. $U=\sum_{i} c_i V_i$, can also be utilized for the same task; however, as expectation values are calculated as $\bra{0}U^\dagger O U\ket{0}$ where $\ket{0}$ is an initial state, this approach requires us to evaluate $\sum_{i,j} c_ic_j^* \bra{0}V_j^\dagger O V_i\ket{0}$ which are rather hard for the NISQ devices.
This fact demonstrates the advantage of using the above formalism.
The tensor network representation of the superoperator formalism allows us to graphically understand the decompositions.

\subsection{Virtual two-qubit gate}
We can show the following, which can then be utilized to decompose any two-qubit gate into a sequence of single-qubit operations.
\begin{lemma}\label{thm:two_to_single}
    For operators $A_1$ and $A_2$ such that $A_1^2=I$ and $A_2^2=I$,
    \begin{align}
        &\mathcal{S}(e^{i\theta A_1\otimes A_2}) = \cos^2\theta \mathcal{S}(I\otimes I) + \sin^2\theta \mathcal{S}(A_1\otimes A_2)+ \nonumber\\
        &\frac{1}{8}\cos\theta\sin\theta\sum_{(\alpha_1,\alpha_2)\in\{\pm 1\}^2} \alpha_1 \alpha_2\left[\mathcal{S}((I+ \alpha_1 A_1)\otimes (I+ i\alpha_2  A_2)) \right. \nonumber\\
        &\qquad\qquad\qquad\qquad\qquad \left.+ \mathcal{S}((I+ i\alpha_1 A_1)\otimes (I+ \alpha_2  A_2))\right]
    \end{align}
\end{lemma}

\begin{figure*}
    \includegraphics[width=\linewidth]{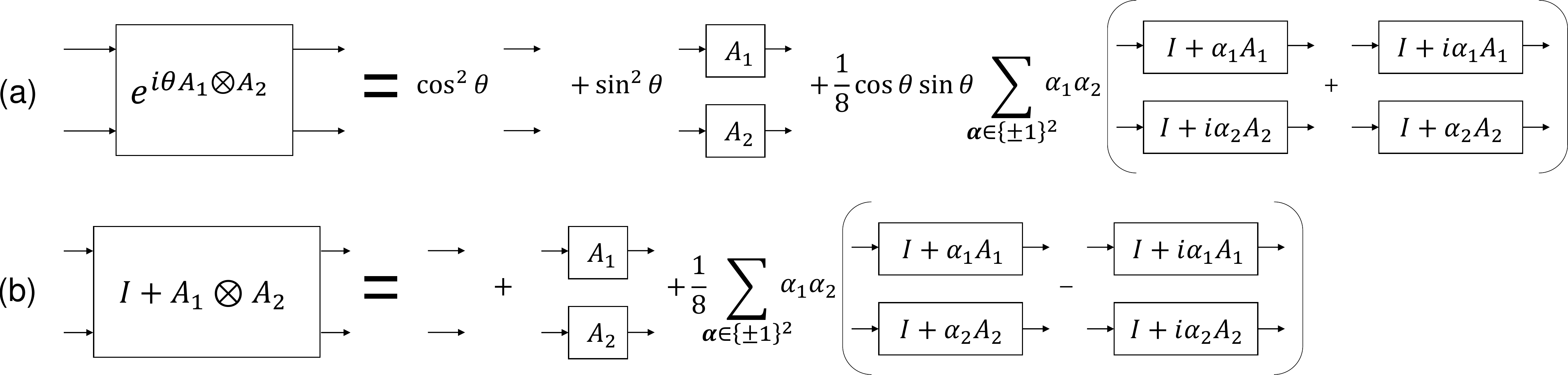}
    \caption{\label{fig:two_qubit_gate_cut} Decomposition of (a) a non-local gate and (b) a non-local non-destructive measurement into a sequence of local operations. $A_1$ and $A_2$ are operators such that $A_1^2 = I$ and $A_2^2 = I$.
    }
\end{figure*}

To prove this, we can directly check the tensor representation of both hand side is equivalent. For detailed calculation, see Appendix \ref{app:two_to_single}.
This theorem is schematically depicted in Fig. \ref{fig:two_qubit_gate_cut} (a).
Notice that the operation that is proportional to $I\pm A$ and $I \pm iA$ for $A \in \{X, Y, Z\}$ can respectively be performed by a projective measurement and a single-qubit rotation.

The correspondence with a single-qubit rotation is clear from the formula, $e^{\pm i\pi A/4} = \frac{1}{\sqrt{2}}(I\pm iA)$, which is the rotation of angle $\pi/2$ around the $A$ axis.
Let $\mathcal{M}_A$ be the projective measurement on the $A$ basis ($A\in \{X,Y,Z\}$), that is, $\mathcal{M}_A$ acts on a density matrix $\rho$ as,
\begin{align}
    \mathcal{M}_A\rho &= \frac{1}{\Tr\left(\rho \frac{I + \alpha A}{2}\right)} \left(\frac{I+\alpha A}{2}\right)\rho \left(\frac{I+\alpha A}{2}\right),
\end{align}
depending on the result of the measurement $\alpha \in \{1,-1\}$.
This is equivalent to $\mathcal{S}(I\pm A)$ up to the factor of  $4\Tr\left(\rho \frac{I + \alpha A}{2}\right)$, that is,
\begin{align}\label{eq:corresp_measurement}
    \mathcal{S}(I + \alpha A) &= 4\Tr\left(\rho \frac{I + \alpha A}{2}\right) \mathcal{M}_{A,\alpha},
\end{align}
where $\mathcal{M}_{A,\alpha}$ is a measurement operation postselected with the measurement outcome $\alpha$.
$\Tr\left(\rho \frac{I + \alpha A}{2}\right)$ is the probability of getting the result $\alpha$ by measuring $\rho$ on the $A$ basis.
Lemma \ref{thm:two_to_single} with this fact implies that the gate $e^{i\theta A_1\otimes A_2}$ can be decomposed, in a sense of Eq.~(\ref{eq:superop_sum_decomposition}), into a sum of $I\otimes I$, $A_1\otimes A_2$, $\mathcal{M}_{A_1}\otimes e^{\pm i\pi A_2/4}$, and $e^{\pm i\pi A_1/4}\otimes \mathcal{M}_{A_2}$, which can be stated as Lemma below.
Notably, this technique can be applied for any $\theta$, which enables us to perform continuous two-qubit gates.
\begin{lemma}\label{thm:two_qubit_gate_decomposition}
    A quantum gate $e^{i\theta A_1\otimes A_2}$ with operators $A_1$ and $A_2$ such that $A_1^2=I$ and $A_2^2=I$ can be decomposed into 6 single-qubit operations.
    For any quantum state $\kket{\rho}$, to achieve the error $\epsilon$ of the decomposition with respect to the trace distance with probability at least $1-\delta$, the required number of circuit runs is $O(\log (1/\delta)/\epsilon^2)$.
\end{lemma}
The detailed proof is given in Appendix \ref{appsec:proof_of_two_qubit_gate_decomp}.
Intuitively, since the error comes from the probabilistic part of the decomposition, that is the renomalization factor in Eq. (\ref{eq:corresp_measurement}) $\Tr\left(\rho \frac{I + \alpha A}{2}\right)$,  if we want to estimate $\Tr\left(\rho \frac{I + \alpha A}{2}\right)$ within error $\epsilon$, $O(1/\epsilon^2)$ repetition would suffice.

\begin{figure*}
    \includegraphics[width=0.8\linewidth]{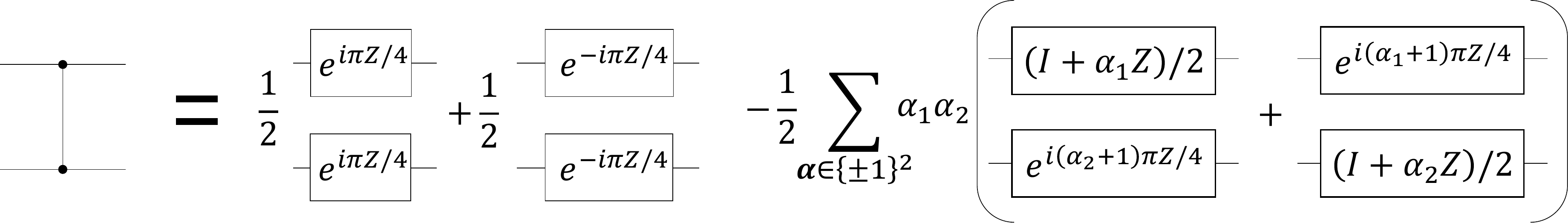}
    \caption{\label{fig:cz_gate_cut} Decomposition of controlled-Z gate into a sequence of single-qubit operations.
    }
\end{figure*}

Let us finally mention the case of the controlled-Z gate, which we denote by $CZ$.
$CZ$ can be decomposed into
\begin{equation}
    CZ = e^{i\pi I\otimes Z/4}e^{i\pi Z\otimes I/4}e^{-i\pi Z\otimes Z/4},
\end{equation}
ignoring the global phase.
This means we can decompose a CZ gate using Lemma \ref{thm:two_qubit_gate_decomposition}.
The decomposition is shown in Fig.~\ref{fig:cz_gate_cut}.
Similar decompositions can be performed on some basic two-qubit gates such as CNOT.
Endo \textit{et al.} \cite{Endo2018} also provides such decomposition (Ref. \cite{Endo2018}, Appendix B).
However, our protocol above is slightly advantageous in that the number of single-qubit operations required is 6 compared to theirs which requires 9 of them.

\subsection{Virtual non-destructive measurement of two-qubit operators}
In the previous subsection, we showed that any two-qubit rotation can be decomposed into a sum of single-qubit operations.
Here, we extend the strategy to construct virtual non-destructive measurement
of two-qubit operators.
Similar to the previous section, we can show the following.
This theorem is schematically shown in Fig.~\ref{fig:two_qubit_gate_cut}~(b).
\begin{lemma}\label{thm:two_to_single_meas}
    For operators $A_1$ and $A_2$ such that $A_1^2=I$ and $A_2^2=I$,
    \begin{align}
        &\mathcal{S}(I + A_1\otimes A_2) = \mathcal{S}(I\otimes I) + \mathcal{S}(A_1\otimes A_2)+ \nonumber\\
        &\frac{1}{8}\sum_{(\alpha_1, \alpha_2)\in\{\pm 1\}^2} \alpha_1 \alpha_2\left[\mathcal{S}((I+ \alpha_1 A_1)\otimes (I+ \alpha_2  A_2)) \right. \nonumber\\
        &\qquad\qquad\qquad \left.- \mathcal{S}((I+ i\alpha_1 A_1)\otimes (I+ i\alpha_2  A_2))\right]
    \end{align}
\end{lemma}
This can also be shown by the direct calculation of both hand side. See Appendix \ref{app:two_to_single_meas} for detailed calculation.

The above Lemma can be utilized to show the following.
\begin{lemma}\label{thm:twoq_meas_decomp}
    A non-local projection $\frac{\red{I}+A_1\otimes A_2}{2}$ with operators $A_1$ and $A_2$ such that $A_1^2=1$ and $A_2^2=2$ can be decomposed into 6 single-qubit operations.
    For any quantum state $\kket{\rho}$, to achieve the error $\epsilon$ of the decomposition with respect to the trace distance with probability at least $1-\delta$, the required number of circuit runs is $O(\log (1/\delta)/\epsilon^2)$.
\end{lemma}
This can be shown with exactly the same approach taken to prove Lemma \ref{thm:two_qubit_gate_decomposition}, which is provided in Appendix \ref{appsec:proof_of_two_qubit_gate_decomp}.

\section{Application}
\subsection{Simulation of large quantum circuits}\label{sec:simulation}
\begin{figure}
    \includegraphics[width=0.7\linewidth]{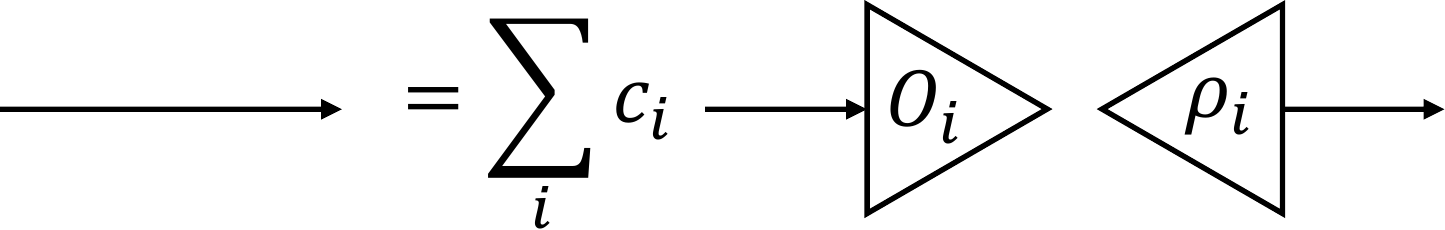}
    \caption{\label{fig:identitygatecut} Time-like cut employed in Ref.~\cite{Peng2019}.}
\end{figure}

The idea of simulating a large quantum circuit by a small quantum computer has been put forward in Ref.~\cite{Peng2019}.
Peng \textit{et al.} utilized the equivalence shown in Fig. \ref{fig:identitygatecut}.
In the figure,
\begin{equation}\label{eq:Pengcut_pairs}
\begin{array}{lll}
    O_1=I, & \rho_1 = \ket{0}\bra{0}, & c_1 = +1/2, \\
    O_2=I, & \rho_2 = \ket{1}\bra{1}, & c_2 = +1/2, \\
    O_3=X, & \rho_3 = \ket{+}\bra{+}, & c_3 = +1/2, \\
    O_4=X, & \rho_4 = \ket{-}\bra{-}, & c_4 = -1/2, \\
    O_5=Y, & \rho_5 = \ket{+i}\bra{+i}, & c_5 = +1/2, \\
    O_6=Y, & \rho_6 = \ket{-i}\bra{-i}, & c_6 = -1/2, \\
    O_7=Z, & \rho_7 = \ket{0}\bra{0}, & c_7 = +1/2, \\
    O_8=Z, & \rho_5 = \ket{1}\bra{1}, & c_8 = -1/2, 
\end{array}
\end{equation}
where $\ket{\pm}=(\ket{0}\pm \ket{1})/\sqrt{2}$ and $\ket{\pm i}=(\ket{0}\pm i\ket{1})/\sqrt{2}$. 
The symbols $\triangleright$ and $\triangleleft$ denotes the measurement of a certain observable and the preparation of a certain state, respectively.
Contrasting this technique and ours, we refer to the former and the latter as ``time-like'' and ``space-like'' cut, respectively.
More concretely, \textit{a time-like cut} of a quantum channel can be defined as a decomposition  of the channel in the sense of Eq. (\ref{eq:superop_sum_decomposition}) using measure-and-prepare channels only.
In contrast, \textit{a space-like cut} of a non-local quantum channel is a decomposition of the channel using local quantum channels only.

\begin{figure}
    \includegraphics[width=\linewidth]{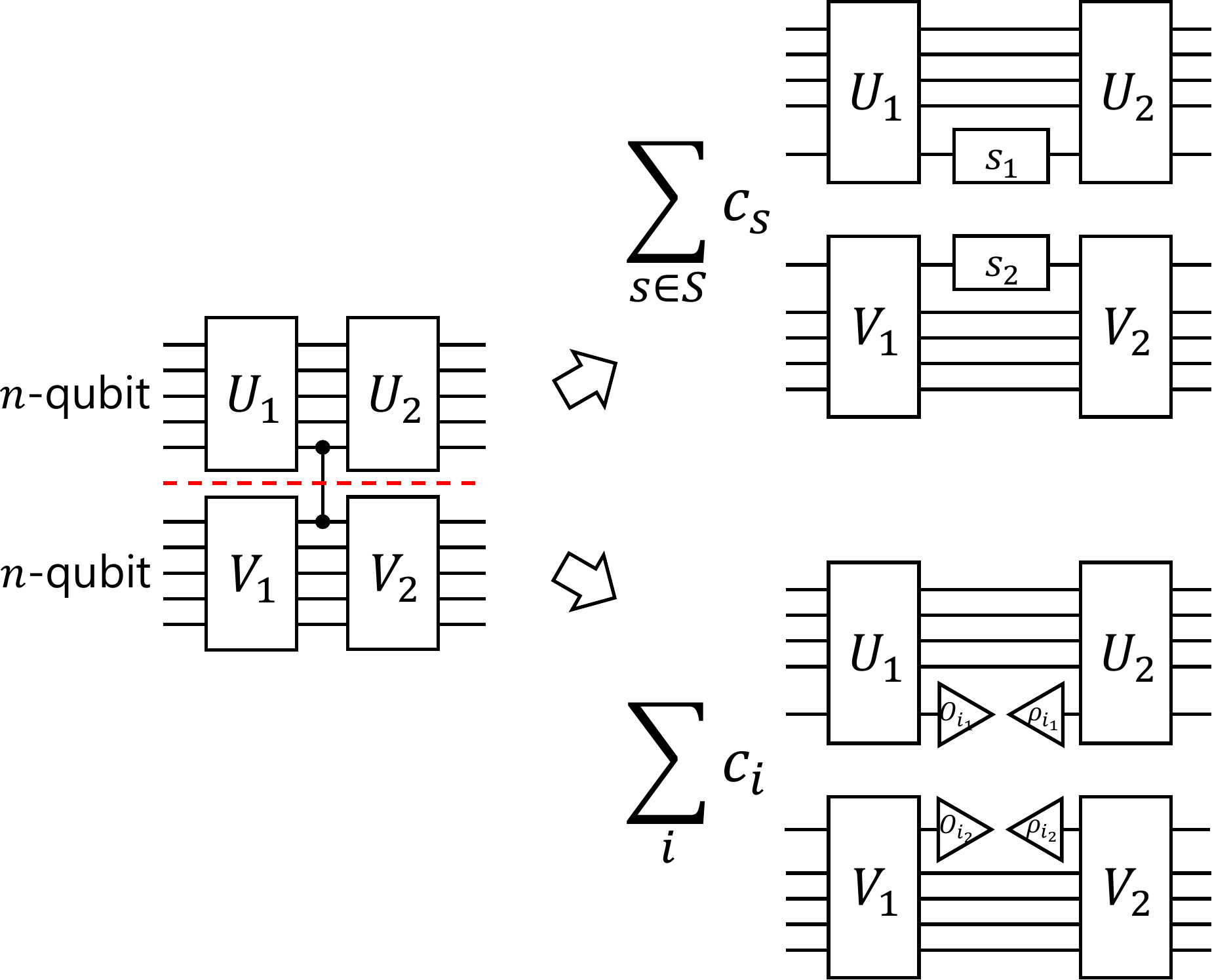}
    \caption{\label{fig:simpleclustering} Two decomposition approach compared in main text. The top-right approach is the presented, and the bottom-right approach is of Ref.~\cite{Peng2019}.}
\end{figure}

The decomposition presented in the previous section can also be used in this direction.
Let us compare the scaling of cost of our decomposition scheme and that of Peng \textit{et al.} by a simple example.
We consider the case where we have an $n$-qubit quantum computer to simulate a $2n$-qubit quantum circuit of Fig.~\ref{fig:simpleclustering}, which has only one CZ gate between n-qubit ``cluster''.
The task is to estimate the expectation value of a final observable $O_f$ by measuring it in the computational basis.
To simplify the discussion, we assume $O_f$ is a string of Pauli $Z$'s.

Let $v$ be a desired variance of the estimation of the expectation value of $O_f$.
We can show a naive algorithm, which runs the equal number of circuits for each terms appearing in the decomposition, to perform the decomposition with time-like cuts, in the worst case, requires $2048/v$ runs of $n$-qubit circuit, while the space-like cut approach takes $\frac{15}{2v}$ runs.
The analysis of this simple example is given in Appendix \ref{app:simpleexample}.
Although the analysis given here is based on a naive algorithm and there are possibilities to improve it, this analysis somewhat shows the enhancement provided by our space-like cut protocol.

\subsubsection*{General case}
We can consider a general case where we perform the time-like and space-like cuts simultaneously to make a given $m$-qubit quantum circuit runnable on an $n$-qubit quantum computer.
Let the number of time-like and space-like cuts be $M_t$ and $M_s$, respectively.
For space-like cuts, we assume they are performed only on CZ gates.
The input state $\rho$ is initialized in $\ket{0}\bra{0}^{\otimes m}$ and $O_f$ is an output (diagonal) observable calculated from some output function $f:\{0,1\}^m\to [-1,1]$.
Our task here is to estimate the expectation $\mathbb{E}[f(y)]$ for a random bitstring $y\in\{0,1\}^m$ sampled from the original circuit.
This model is adopted from Ref. \cite{Peng2019} which originates in Ref. \cite{Bravyi2016}.
With this definition, we can get the following.

\begin{theorem}\label{thm:multiple_twoqgate_cut}
    The number of $n$-qubit circuit runs required to estimate $\mathbb{E}[f(y)]$ within accuracy $\epsilon$ with some high probability $1-\delta$ is $O\left(\frac{9^{M_s} 16^{M_t}}{\epsilon^2}\log\left(\frac{1}{2\delta}\right)\right)$.
\end{theorem}
This implies that the decomposition of the circuit should be performed to minimize $9^{M_s} 16^{M_t}$.
A detailed proof is given in Appendix \ref{appsec:proof_of_multiple_twoqgate_cut}, however, the above can roughly be explained as follows.
At each space-like cut, we get 6 different sets of single-qubit operations, so $M_s$ cuts induce $6^{M_s}$ terms.
Likewise, $M_t$ time-like cuts induce $8^{M_t}$ terms, which makes the total number of circuits in decomposition $6^{M_s}8^{M_t}$.
With this decomposition, we can take a Monte-Carlo approach to estimate the sum, that is, we randomly choose circuits to run and average them.
Hoeffding's inequality can be used to bound the error of such protocol, which states that if a magnitude of a random variable is always bounded by some constant $a$, then $O(a^2/\epsilon^2)$ samples would suffice to obtain an accuracy of $\epsilon$.
In this case, we are to estimate $\mathbb{E}[f(y)] = \sum_{i=1}^{6^{M_s}8^{M_t}} c_i \bbra{O_f}\mathcal{S}(V_i)\kket{\rho}$ with $i$ randomly drawn from $\{1,\cdots,6^{M_s}8^{M_t}\}$ and $|c_i|=1/2^{M_s+M_t}$, that is, $\mathbb{E}[f(y)]$ is estimated by $\mathbb{E}_i[6^{M_s}8^{M_t} c_i\bbra{O_f}\mathcal{S}(V_i)\kket{\rho}]$.
The magnitude of random variable $6^{M_s}8^{M_t} c_i\bbra{O_f}\mathcal{S}(V_i)\kket{\rho}$ is roughly $3^{M_s}4^{M_t}$, thus we can apply the Hoeffding bound to get the result.

\begin{figure}
    \includegraphics[width=\linewidth]{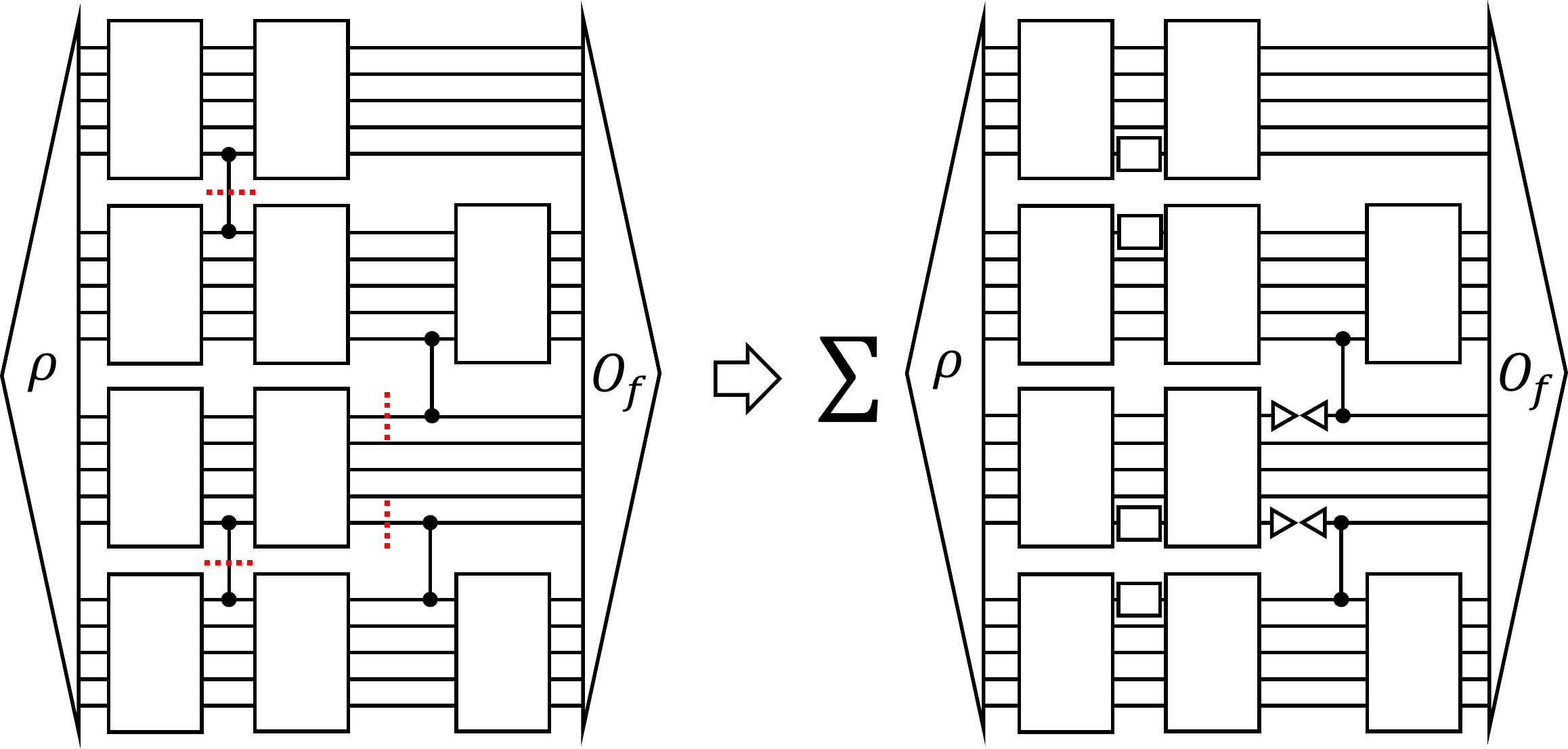}
    \caption{\label{fig:combination}Schematic illustration of performing the space-like cut and the time-like cut simultaneously.}
\end{figure}

\subsection{Distant two-qubit gates}
\begin{figure}
    \includegraphics[width=\linewidth]{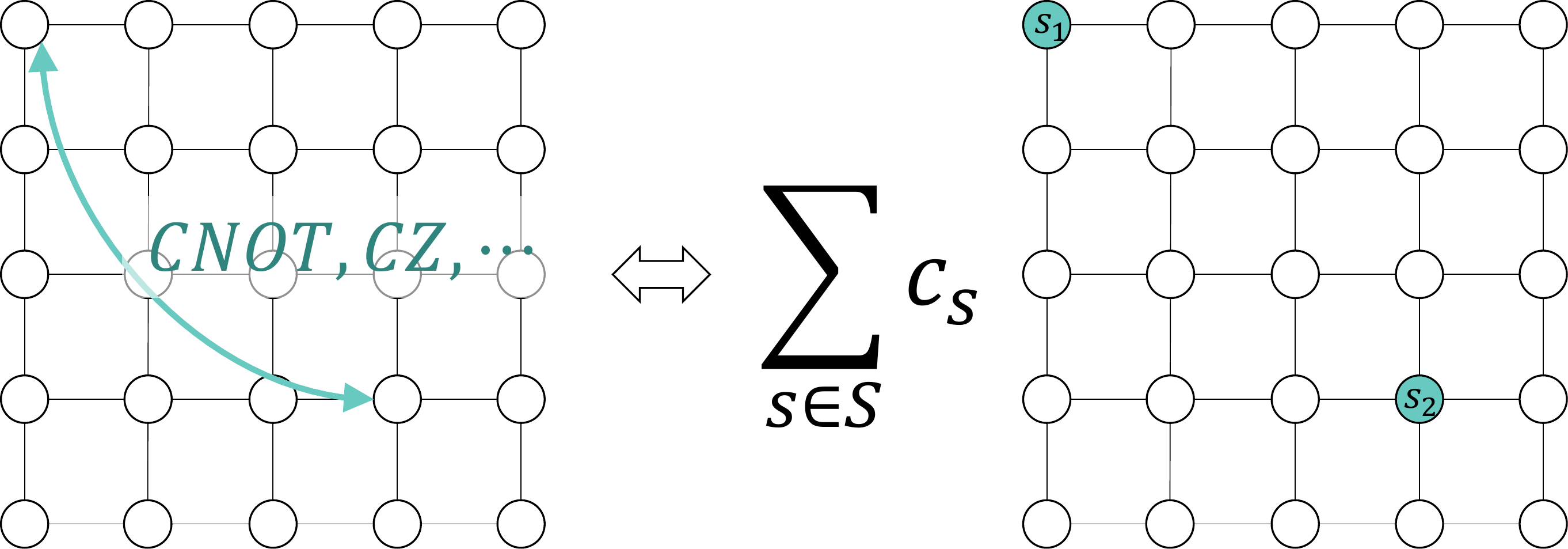}
    \caption{\label{fig:distant_2qubit_gate} Decomposition of distant two-qubit gate on a square lattice. 
    Each vertex of the graph represents a qubit and the edge represents the connectivity of the qubits.
    $S$ is the set of pairs of single-qubit operations which appears in the formula in Lemma \ref{thm:two_to_single}, and $c_s$ is the corresponding coefficient for each pair.}
\end{figure}

The theorem introduced above can be utilized to ``virtually'' perform a two-qubit gate between qubits at distance.
Figure \ref{fig:distant_2qubit_gate} shows an example of such a virtual two-qubit gate.
Notice that this protocol works irrespective of the distance between the qubits.
Many swap gates are otherwise necessary for performing such gates, which makes them impractical on NISQ devices due to the non-negligible amount of decoherence and gate error of such devices.

This protocol might be useful for the variational algorithms such as the variational quantum eigensolver (VQE) \cite{Peruzzo2014} and the quantum approximate optimization algorithms (QAOA) \cite{Farhi2014}.
Here, we describe an example in the QAOA.
In the QAOA, we seek to find a ground state of a Hamiltonian $H$ on $n$-qubit which is a sum of Pauli $Z$'s and its products.
For example, a Hamiltonian may have the form of,
\begin{align}
    H = \sum_{ij} J_{ij} Z_iZ_j.
\end{align}
The QAOA tries to solve the problem by converting it to a optimization problem of a continuous variable $\bm{\beta}$ and $\bm{\gamma}$.
The optimization of $\bm{\beta}$ and $\bm{\gamma}$ are performed so as to minimize the function,
\begin{align}
    \expect{H(\bm{\beta}, \bm{\gamma})} =\bra{+}^{\otimes n} U^\dagger(\bm{\beta}, \bm{\gamma}) H U(\bm{\beta}, \bm{\gamma})\ket{+}^{\otimes n},
\end{align}
where,
\begin{align}\label{eq:qaoa-circuit}
    U(\bm{\beta}, \bm{\gamma}) = e^{i\beta_p \sum_i X_i}e^{i\gamma_p H} \cdots e^{i\gamma_2 H}e^{i\beta_1 \sum_i X_i}e^{i\gamma_1 H}.
\end{align}
This algorithm has been experimentally demonstrated \cite{Otterbach2017} with the connectivity of the target Hamiltonian being equivalent to the connectivity of the actual device.

The equivalence of the connectivity is almost necessary from the requirement to perform $e^{i\gamma H}$.
This requirement can somewhat be relaxed by our protocol which enables qubits to virtually interact irrespective of the distance between them.
Let us now assume that an available device has a square-lattice connectivity of Fig. \ref{fig:distant_2qubit_gate}, and a Hamiltonian of the QAOA which we aim to solve has a interaction between one pair of qubits that is not included in the hardware connectivity graph.
In this case, to execute the QAOA circuit (Eq. (\ref{eq:qaoa-circuit})), we can use our space-like technique $p$ times to virtually apply the unitary.
The scaling of the cost can be bounded by setting $M_t=0$ and $M_s=p$ in Theorem \ref{thm:multiple_twoqgate_cut} which gives us a scaling of $9^p\epsilon^{-2}\log[1/(2\delta)]$.
The time-like cut approach of Peng \textit{et al.} \cite{Peng2019} can also be utilized in this direction.
However, as this approach would require 4 cuts per gate, the cost scaling is bounded by $16^{4p}\epsilon^{-2}\log[1/(2\delta)]$ by setting $M_t=4p$ and $M_s=0$ in Theorem \ref{thm:multiple_twoqgate_cut}.
This demonstrates an advantage, albeit in this special settings, of our technique over the previous result.

In the context of the VQE, which is also an algorithm to find a ground state of a Hamiltonian but mainly targets a concrete physical system such as molecules, it has been proposed to use the same kind of quantum circuits as the QAOA \cite{Wecker2015, Mitarai2019g}.
Our result may also be applicable in constructing such circuits.

\section{discussion and conclusion}
We described a technique to decompose a non-local operations into a sequence of local operations.
\red{As the single-qubit operations are generally more accurate on NISQ devices, the proposed technique can be used to enhance their capability. We believe intrinsic noise on single-qubit operations can be compensated by recent sophisticated error mitigation techniques \cite{Endo2018}.}
In particular, our technique of the space-like cut of two-qubit gates can improve the simulation of a large quantum circuit with a small quantum computer in some cases.
It would be interesting to investigate the best strategy to perform ``cuts'' to reduce the number of qubits compatible with an available device.
Also, the algorithm we have given to bound the cost scaling is rather straight forward and we believe it can be improved with a more sophisticated strategy.

\red{The proposed algorithm can also be compared to the classical simulation strategy that splits a large circuit by decomposing two-qubit gates. For example, a controlled-NOT gate can be splitted using a tensor network based technique \cite{biamonte2017tensor}. However, such techniques generally does not focus on decompositions of $\mathcal{S}(U)$ considered in this work but rather the two-qubit unitary $U$ itself, which takes makes them difficult to be used on NISQ devices as Eq. (\ref{eq:superop_sum_decomposition}) cannot be utilized anymore. }

Our technique can induce a entanglement-like effect without performing any two-qubit gate with the cost mentioned in Lemmas \ref{thm:two_qubit_gate_decomposition} and \ref{thm:twoq_meas_decomp}.
This connects this work to areas like quantum communication.
This ``virtual'' entanglement creation could be done with the time-like cut proposed by Peng \textit{et al.}, but our work lowered the cost to perform the task.
It is interesting to know whether ours is the optimal protocol or there is a more efficient way.

To summarize, our technique allows qubits to virtually interact irrespective of physical distances between them.
The result is useful for applying a two-qubit gate to a distant pair of qubits.
In particular, when applied to the NISQ devices, this may be employed to enhance the power of them.
Future direction can be to explore if we can lower the resource to perform such virtual operations.

\begin{acknowledgements}
    KM thanks the METI and IPA for their support through the MITOU Target program.
    KM is also supported by JSPS KAKENHI No. 19J10978 and No. 20K22330, and JST PRESTO JPMJPR2019.
	KF is supported by KAKENHI No.16H02211, JST PRESTO JPMJPR1668, JST ERATO JPMJER1601, and JST CREST JPMJCR1673.
	The authors thank Suguru Endo for fruitful discussions and letting us become aware of Ref.  \cite{Endo2018}.
    This work is supported by MEXT Quantum Leap Flagship Program (MEXT Q-LEAP) Grant Number JPMXS0118067394.
\end{acknowledgements}

\appendix
\begin{widetext}
\section{Proof of Lemmas \ref{thm:two_to_single} and \ref{thm:two_to_single_meas}}\label{app:theorem1}
A tensor representation of $\mathcal{S}((I+ \alpha_1 A_1)\otimes (I+ \alpha_2 A_2))$ on a set of basis $\{\kket{e_i e_j}\}_{i,j=1}^4$ is as follows.
    \begin{align}
        &\bbra{e_i e_j} \mathcal{S}(I+ \alpha_1 A_1) \otimes (I+ \alpha_2 A_2)\kket{e_k e_l}\nonumber\\
        &= \Tr\left(e_i\otimes e_j(I+\alpha_1 A_1)\otimes (I+\alpha_2 A_2) e_k\otimes e_l (I+\alpha_1^* A_1)\otimes (I+\alpha_2^* A_2)\right)\nonumber\\
        &= \Tr\left(e_i\otimes e_j~e_k\otimes e_l \right)\nonumber\\
        &\quad +\alpha_1\Tr\left(e_i\otimes e_j (A_1\otimes I) e_k\otimes e_l\right) + \alpha_1^*\Tr\left(e_k\otimes e_l(A_1\otimes I)e_i\otimes e_j\right)\nonumber\\
        &\quad +\alpha_2\Tr\left(e_i\otimes e_j (I\otimes A_2) e_k\otimes e_l\right) + \alpha_2^*\Tr\left(e_k\otimes e_l(I\otimes A_2)e_i\otimes e_j\right)\nonumber\\
        &\quad +\alpha_1\alpha_2\Tr\left(e_i\otimes e_j (A_1\otimes A_2) e_k\otimes e_l\right) + \alpha_1^*\alpha_2^*\Tr\left(e_k\otimes e_l(A_1\otimes A_2)e_i\otimes e_j\right)\nonumber\\
        &\quad +\alpha_1\alpha_2^*\Tr\left(e_i\otimes e_j (A_1\otimes I) e_k\otimes e_l(I\otimes A_2)\right) + \alpha_1^*\alpha_2\Tr\left(e_i\otimes e_j(I\otimes A_2)e_k\otimes e_l (A_1\otimes I)\right)\\
        &\quad +\alpha_1\Tr\left(e_i\otimes e_j (A_1\otimes A_2) e_k\otimes e_l(I\otimes A_2)\right) + \alpha_1^*\Tr\left(e_i\otimes e_j(I\otimes A_2)e_k\otimes e_l (A_1\otimes A_2)\right)\nonumber\\
        &\quad +\alpha_2\Tr\left(e_i\otimes e_j (A_1\otimes A_2) e_k\otimes e_l(A_1\otimes I)\right) + \alpha_2^*\Tr\left(e_i\otimes e_j(A_1\otimes I)e_k\otimes e_l (A_1\otimes A_2)\right)\nonumber\\
        &\quad +\Tr\left(e_i\otimes e_j (A_1\otimes A_2) e_k\otimes e_l(A_1\otimes I)\right) + \Tr\left(e_i\otimes e_j(A_1\otimes I)e_k\otimes e_l (A_1\otimes A_2)\right)\nonumber\\
        &\quad +\Tr\left(e_i\otimes e_j (A_1\otimes A_2) e_k\otimes e_l (A_1\otimes A_2)\right)\nonumber.
    \end{align}
    Let,
    \begin{align}
        \{\alpha_1, \alpha_2\}_{ij,kl} &:= \mathcal{S}((I+\alpha_1 A_1)\otimes(I+\alpha_2 A_2))_{ij,kl},\\
        \left(\begin{array}{c}
            a_{1,ijkl}\\
            a_{2,ijkl}\\
            a_{3,ijkl}\\
            a_{4,ijkl}\\
            a_{5,ijkl}\\
            a_{6,ijkl}\\
            a_{7,ijkl}\\
            a_{8,ijkl}\\
            a_{9,ijkl}\\
            a_{10,ijkl}\\
            a_{11,ijkl}\\
            a_{12,ijkl}\\
            a_{13,ijkl}\\
            a_{14,ijkl}\\
            a_{15,ijkl}\\
            a_{16,ijkl}\\
        \end{array}\right)&:=
        \left(\begin{array}{c}
            \Tr\left(e_i\otimes e_j~e_k\otimes e_l \right) \\
            \Tr\left(e_i\otimes e_j (A_1\otimes I) e_k\otimes e_l\right) \\
            \Tr\left(e_k\otimes e_l (A_1\otimes I) e_i\otimes e_j\right) \\
            \Tr\left(e_i\otimes e_j (I\otimes A_2) e_k\otimes e_l\right) \\
            \Tr\left(e_k\otimes e_l (I\otimes A_2) e_i\otimes e_j\right) \\
            \Tr\left(e_i\otimes e_j (A_1\otimes A_2) e_k\otimes e_l\right) \\
            \Tr\left(e_k\otimes e_l (A_1\otimes A_2) e_i\otimes e_j\right) \\
            \left(e_i\otimes e_j (A_1\otimes I) e_k\otimes e_l(I\otimes A_2)\right) \\
            \left(e_i\otimes e_j (I\otimes A_2) e_k\otimes e_l(A_1\otimes I)\right) \\
            \left(e_i\otimes e_j (A_1\otimes A_2) e_k\otimes e_l(I\otimes A_2)\right) \\
            \left(e_i\otimes e_j (I\otimes A_2) e_k\otimes e_l(A_1\otimes A_2)\right) \\
            \left(e_i\otimes e_j (A_1\otimes A_2) e_k\otimes e_l(A_1\otimes I)\right) \\
            \left(e_i\otimes e_j (A_1\otimes I) e_k\otimes e_l(A_1\otimes A_2)\right) \\
            \left(e_i\otimes e_j (A_1\otimes I) e_k\otimes e_l(A_1\otimes I)\right) \\
            \left(e_i\otimes e_j (I\otimes A_2) e_k\otimes e_l(I\otimes A_2)\right) \\
            \left(e_i\otimes e_j (A_1\otimes A_2) e_k\otimes e_l(A_1\otimes A_2)\right) \\
        \end{array}\right).
    \end{align}
    The relation can be summarized in matrix form,
    \begin{align}
        &\left(\begin{array}{c}
            \{+1, +1\}_{ij,kl}\\
            \{+1, +i\}_{ij,kl}\\
            \{+1, -1\}_{ij,kl}\\
            \{+1, -i\}_{ij,kl}\\
            \{+i, +1\}_{ij,kl}\\
            \{+i, +i\}_{ij,kl}\\
            \{+i, -1\}_{ij,kl}\\
            \{+i, -i\}_{ij,kl}\\
            \{-1, +1\}_{ij,kl}\\
            \{-1, +i\}_{ij,kl}\\
            \{-1, -1\}_{ij,kl}\\
            \{-1, -i\}_{ij,kl}\\
            \{-i, +1\}_{ij,kl}\\
            \{-i, +i\}_{ij,kl}\\
            \{-i, -1\}_{ij,kl}\\
            \{-i, -i\}_{ij,kl}
        \end{array}\right)
        &= \left(\begin{array}{cccccccccccccccc}
            1 & 1 & 1 & 1 & 1 & 1 & 1 & 1 & 1 & 1 & 1 & 1 & 1 & 1 & 1 & 1\\
            1 & 1 & 1 & i & -i& i & -i& -i& i & 1 & 1 & i & -i& 1 & 1 & 1\\
            1 & 1 & 1 & -1& -1& -1& -1& -1& -1& 1 & 1 & -1& -1& 1 & 1 & 1\\
            1 & 1 & 1 & -i& i & -i& i & i & -i& 1 & 1 & -i& i & 1 & 1 & 1\\
            1 & i & -i& 1 & 1 & i & -i& i & -i& i & -i& 1 & 1 & 1 & 1 & 1\\
            1 & i & -i& i & -i& -1& -1& 1 & 1 & i & -i& i & -i& 1 & 1 & 1\\
            1 & i & -i& -1& -1& -i& i & -i& i & i & -i& -1& -1& 1 & 1 & 1\\
            1 & i & -i& -i& i & 1 & 1 & -1& -1& i & -i& -i& i & 1 & 1 & 1\\
            1 & -1& -1& 1 & 1 & -1& -1& -1& -1& -1& -1& 1 & 1 & 1 & 1 & 1\\
            1 & -1& -1& i & -i& -i& i & i & -i& -1& -1& i & -i& 1 & 1 & 1\\
            1 & -1& -1& -1& -1& 1 & 1 & 1 & 1 & -1& -1& -1& -1& 1 & 1 & 1\\
            1 & -1& -1& -i& i & i & -i& -i& i & -1& -1& -i& i & 1 & 1 & 1\\
            1 & -i& i & 1 & 1 & -i& i & -i& i & -i& i & 1 & 1 & 1 & 1 & 1\\
            1 & -i& i & i & -i& 1 & 1 & -1& -1& -i& i & i & -i& 1 & 1 & 1\\
            1 & -i& i & -1& -1& i & -i& i & -i& -i& i & -1& -1& 1 & 1 & 1\\
            1 & -i& i & -i& i & -1& -1& 1 & 1 & -i& i & -i& i & 1 & 1 & 1
        \end{array}\right)
        \left(\begin{array}{c}
            a_1\\
            a_2\\
            a_3\\
            a_4\\
            a_5\\
            a_6\\
            a_7\\
            a_8\\
            a_9\\
            a_{10}\\
            a_{11}\\
            a_{12}\\
            a_{13}\\
            a_{14}\\
            a_{15}\\
            a_{16}\\
        \end{array}\right).
    \end{align}
    \subsection{Proof of Lemma \ref{thm:two_to_single}}\label{app:two_to_single}
    Tensor representation of $\mathcal{S}(e^{i\theta A_1\otimes A_2})$ is,
    \begin{align}\label{eq:two_qubit_gate_tensor}
        &\bbra{e_i e_j} \mathcal{S}(e^{i\theta A_1\otimes A_2})\kket{e_k e_l}\nonumber\\
        &= \bbra{e_i e_j} \mathcal{S}(\cos\theta + i\sin\theta A_1\otimes A_2)\kket{e_k e_l} \nonumber \\
        &= \Tr\left(e_i\otimes e_j (\cos\theta I+i\sin\theta A_1\otimes A_2) e_k\otimes e_l (\cos\theta I-i\sin\theta A_1\otimes A_2)\right) \nonumber\\
        &= \cos^2\theta a_{1,ijkl} +i\sin \theta \cos \theta \left(a_{6,ijkl} - a_{7,ijkl}\right)+\sin^2\theta~a_{16,ijkl}.
    \end{align}
\end{widetext}
Observe that,
\begin{equation}
    \left\{
\begin{array}{l}
    \{+1,+i\}-\{+1,-i\} = \\
    2i(a_{4}-a_{5}) + 2i(a_{6}-a_{7}) - 2i(a_{8}-a_{9}) + 2i(a_{12}-a_{13}), \\
    \{-1,+i\}-\{-1,-i\} = \\
    2i(a_{4}-a_{5}) - 2i(a_{6}-a_{7}) + 2i(a_{8}-a_{9}) + 2i(a_{12}-a_{13}), \\
    \{+i,+1\}-\{-i,+1\} = \\
    2i(a_{2}-a_{3}) + 2i(a_{6}-a_{7}) + 2i(a_{8}-a_{9}) + 2i(a_{12}-a_{13}),\\
    \{+i,-1\}-\{-i,-1\} = \\
    2i(a_{2}-a_{3}) - 2i(a_{6}-a_{7}) - 2i(a_{8}-a_{9}) + 2i(a_{12}-a_{13}),
\end{array}\right.
\end{equation}
where we abbreviated the subscripts $ijkl$.
We can solve the above for $i(a_{6}-a_{7})$, and obtain
\begin{align}
    &8i(a_{6}-a_{7})\nonumber\\
    &=\{+1,+i\} - \{+1,-i\} - \{-1,+i\} + \{-1,-i\}\nonumber\\
    &\quad +\{+i,+1\}-\{-i,+1\} - \{+i,-1\}+\{-i,-1\}\\
    &=\sum_{\bm{\alpha}\in\{\pm 1\}^2} \alpha_1 \alpha_2\left[\mathcal{S}((I+ \alpha_1 A_1)\otimes (I+ i\alpha_2  A_2)) \right. \nonumber\\
    &\qquad\qquad\qquad \left.+ \mathcal{S}((I+ i\alpha_1 A_1)\otimes (I+ \alpha_2  A_2))\right].
\end{align}
Combining this with Eq. (\ref{eq:two_qubit_gate_tensor}) completes the proof.

\subsection{Proof of Lemma \ref{thm:two_to_single_meas}}\label{app:two_to_single_meas}
We first write down the tensor representation of the projective measurement, $I+\beta A_1\otimes A_2$ for $\beta = \pm 1$.
\begin{align}\label{eq:two_qubit_measurement_tensor}
    &\bbra{e_i e_j} \mathcal{S}(I+\beta A_1\otimes A_2)\kket{e_k e_l}\nonumber\\
    &= \Tr\left(e_i\otimes e_j (I+\beta A_1\otimes A_2) e_k\otimes e_l ( I+\beta A_1\otimes A_2)\right) \nonumber\\
    &= a_{1,ijkl} + \beta \left(a_{6,ijkl} + a_{7,ijkl}\right)+a_{16,ijkl}
\end{align}
Similarly to the previous proof, observe that,
\begin{equation}
    \left\{
\begin{array}{l}
    \{+1,+1\}-\{+1,-1\} = \\
    2(a_{4}+a_{5}) + 2(a_{6}+a_{7}) + 2(a_{8}+a_{9}) + 2(a_{12}+a_{13}), \\
    \{-1,+1\}-\{-1,-1\} = \\
    2(a_{4}+a_{5}) - 2(a_{6}+a_{7}) - 2(a_{8}+a_{9}) + 2(a_{12}+a_{13}), \\
    \{+i,+i\}-\{+i,-i\} = \\
    2i(a_{4}-a_{5}) - 2(a_{6}+a_{7}) + 2(a_{8}+a_{9}) + 2i(a_{12}-a_{13}),\\
    \{-i,+i\}-\{-i,-i\} = \\
    2i(a_{4}-a_{5}) + 2(a_{6}+a_{7}) - 2(a_{8}+a_{9}) + 2i(a_{12}-a_{13}),
\end{array}\right.
\end{equation}
We can solve the above for $i(a_{6}+a_{7})$, and obtain,
\begin{align}
    &8(a_{6}+a_{7})\nonumber\\
    &=\{+1,+1\} - \{+1,-1\} - \{-1,+1\} + \{-1,-1\}\nonumber\\
    &\quad -\{+i,+i\}+\{+i,-i\} + \{-i,+i\}-\{-i,-i\}\\
    &=\sum_{\bm{\alpha}\in\{\pm 1\}^2} \alpha_1 \alpha_2\left[\mathcal{S}((I+ \alpha_1 A_1)\otimes (I+ \alpha_2  A_2)) \right. \nonumber\\
    &\qquad\qquad\qquad \left.- \mathcal{S}((I+ i\alpha_1 A_1)\otimes (I+ i\alpha_2  A_2))\right].
\end{align}

\subsection{Relation with Ref. \cite{Bravyi2016}}
Bravyi \textit{et al.} has considered to remove $k$ qubits in a given $n+k$-qubit circuit at the cost of $O(kd 2^{k})$ classical computation, where $d$ defined to be proportinal to the number of gates applied to the $k$-qubit system.
The technique utilized in their work, in particular, Fig. 2 in Ref. \cite{Bravyi2016} can also provide a derivation to the above Lemmas when combined with our recent technique developed in Ref. \cite{Mitarai2019g}.

\section{Proof of Lemma \ref{thm:two_qubit_gate_decomposition}} \label{appsec:proof_of_two_qubit_gate_decomp}
Suppose that we are applying $\mathcal{S}(e^{i\theta A_1\otimes A_2})$ to some state $\rho$ and want to decompose the gate.
We name each operation in the decomposition as,
\begin{align}
    \Phi_{1, \beta} &= \mathcal{S}(I\otimes I), \nonumber \\
    \Phi_{2, \beta} &= \mathcal{S}(A_1\otimes A_2), \nonumber \\
    \Phi_{3,\beta} &= \beta \mathcal{M}_{A_1, \beta} \otimes \mathcal{S}(e^{i\pi A_2/4}), \nonumber\\
    \Phi_{4,\beta} &= \beta \mathcal{M}_{A_1, \beta} \otimes \mathcal{S}(e^{-i\pi A_2/4}), \\
    \Phi_{5,\beta} &= \beta \mathcal{S}(e^{i\pi A_1/4}) \otimes \mathcal{M}_{A_2, \beta}, \nonumber \\
    \Phi_{6, \beta} &= \beta \mathcal{S}(e^{-i\pi A_1/4}) \otimes \mathcal{M}_{A_2, \beta}. \nonumber
\end{align}
which is not physical when $\beta_{3,4,5,6}=-1$ but achivable with classical post processing.
$\mathcal{M}_{A_i, \beta}$ is a postselective measurement operation, which has been introduced in the main text.
For convenience, we define coefficients $\{c_i\}_{i=1}^6$ as
\begin{align}\label{eq:space-like-cut-coefs}
    c_1 &= \cos^2 \theta, \nonumber \\
    c_2 &= \sin^2 \theta, \\
    c_3 &= -c_4 = c_5 = -c_6 = \cos\theta \sin\theta, \nonumber
\end{align}
Then,
\begin{align}
    &\mathcal{S}(e^{i\theta A_1\otimes A_2})\kket{\rho} \\
    &= \left[c_1 \Phi_{1, \beta} + c_2 \Phi_{2, \beta} \right.\\
    &\quad  \sum_{\beta\in\{1,-1\}}\Tr\left(\rho \frac{I + \beta A_1}{2}\right)(c_3\Phi_{3, \beta}+c_4\Phi_{4, \beta}) \\
    &\quad \left.\sum_{\beta\in\{1,-1\}}\Tr\left(\rho \frac{I + \beta A_2}{2}\right)(c_5\Phi_{5, \beta}+c_6\Phi_{6, \beta})\right]\kket{\rho}
\end{align}

We take a naive algorithm to bound the error of the decomposition.
We define a probabilistic map below
\begin{align}
    \Phi_{1} &= \mathcal{S}(I\otimes I), \nonumber \\
    \Phi_{2} &= \mathcal{S}(A_1\otimes A_2), \nonumber \\
    \Phi_{3} &= \mathcal{M}_{A_1}' \otimes \mathcal{S}(e^{i\pi A_2/4}), \nonumber\\
    \Phi_{4} &= \mathcal{M}_{A_1}' \otimes \mathcal{S}(e^{-i\pi A_2/4}), \\
    \Phi_{5} &= \mathcal{S}(e^{i\pi A_1/4}) \otimes \mathcal{M}_{A_2}', \nonumber \\
    \Phi_{6} &= \mathcal{S}(e^{-i\pi A_1/4}) \otimes \mathcal{M}_{A_2}', \nonumber
\end{align}
where $\mathcal{M}_{A_i}'$ acts on a state $\rho$ probabilistically as,
\begin{align}
    \mathcal{M}_{A_i}'(\rho) \to b \mathcal{M}_{A_i, b}(\rho)
\end{align}
where $b$ is a random variable with probability distribution $p(b=\pm 1)=\Tr\left(\rho \frac{I \pm A_i}{2}\right)$.
Again, when $b=-1$ this map is non-physical but can be realized with classical post processing.
$\Phi_{i}$ becomes $\Phi_{i,b}$ with probability $\Tr\left(\rho \frac{I \pm A_i}{2}\right)$, and threfore,
\begin{equation}
    \mathbb{E}[\Phi_{i}] = \Tr\left(\rho \frac{I + A_1}{2}\right) \Phi_{i, +1} + \Tr\left(\rho \frac{I - A_1}{2}\right) \Phi_{i, -1}
\end{equation}
for $i=3,4$.
A similar equality holds for $i=5,6$.
This yields,
\begin{align}\label{eq:space-like-cut-decomposition-with-prob-map}
    \mathcal{S}(e^{i\theta A_1\otimes A_2})\kket{\rho} 
    &= \sum_{i=1}^6 c_i \mathbb{E}[\Phi_i\kket{\rho}]
\end{align}

Suppose that we take $N$ samples for each $i=1,\cdots,6$ to estimate $\mathbb{E}[\Phi_i\kket{\rho}]$.
The $i=1,2$ cases are not probabilistic and hence do not introduce error.
We are left to consider the error induced by $i=3,4,5,6$.
In this case, we can estimate $\kket{\mu_i} = \mathbb{E}[\Phi_i\kket{\rho}]$ by
\begin{align}
    \kket{\bar{\mu_i}} = \frac{1}{N} \sum_{j=1}^{N} \Phi_{i, b_{ij}}\kket{\rho}
\end{align}
where $\{b_{ij}\}_{j=1}^N$ are samples drawn from the distribution which is identical to the above mentioned $b$.
Now the difference between the true state $\mathcal{S}(e^{i\theta A_1\otimes A_2})\kket{\rho}$ and the estimated $\sum_{i=1}^6 c_i \kket{\bar{\mu_i}}$ is,
\begin{widetext}
    \begin{align}\label{eq:difference_rhoandmu}
        &\mathcal{S}(e^{i\theta A_1\otimes A_2})\kket{\rho} - \sum_{i=1}^6 c_i \kket{\bar{\mu_i}} = \nonumber \\
        &\left(\Tr\left(\rho \frac{I + A_1}{2}\right)-\frac{1}{2N}\sum_{j=1}^N(b_{3j}+1)\right)c_3\Phi_{3, +1}\kket{\rho} + \left(\Tr\left(\rho \frac{I - A_1}{2}\right)-\frac{1}{2N}\sum_{j=1}^N(1-b_{3j})\right)c_3\Phi_{3, -1}\kket{\rho} \nonumber \\
        &+ \left(\Tr\left(\rho \frac{I + A_1}{2}\right)-\frac{1}{2N}\sum_{j=1}^N(b_{4j}+1)\right)c_4\Phi_{4, +1}\kket{\rho} 
        + \left(\Tr\left(\rho \frac{I - A_1}{2}\right)-\frac{1}{2N}\sum_{j=1}^N(1-b_{4j})\right)c_4\Phi_{4, +1}\kket{\rho} \nonumber \\
        &+ \left(\Tr\left(\rho \frac{I + A_2}{2}\right)-\frac{1}{2N}\sum_{j=1}^N(b_{5j}+1)\right)c_5\Phi_{5, +1}\kket{\rho} 
        + \left(\Tr\left(\rho \frac{I - A_2}{2}\right)-\frac{1}{2N}\sum_{j=1}^N(1-b_{5j})\right)c_5\Phi_{5, -1} \kket{\rho}\nonumber \\
        &+ \left(\Tr\left(\rho \frac{I + A_2}{2}\right)-\frac{1}{2N}\sum_{j=1}^N(b_{6j}+1)\right)c_6\Phi_{6, +1}\kket{\rho} 
        + \left(\Tr\left(\rho \frac{I - A_2}{2}\right)-\frac{1}{2N}\sum_{j=1}^N(1-b_{5j})\right)c_6\Phi_{6, -1} \kket{\rho}.
    \end{align}
\end{widetext}
$\frac{1\pm b_{ij}}{2}$ is a Bernouilli random variable with the expectation $\frac{I \pm A_1}{2}$ and $\frac{I \pm A_2}{2}$ respectively for $i=3,4$ and $i=5,6$.
This means that, for example, the difference between $\frac{1}{2N}\sum_{j=1}^N(b_{3j}+1)$ and $\frac{I + A_1}{2}$ is bounded by $\epsilon>0$, that is, $\left|\frac{1}{2N}\sum_{j=1}^N(b_{3j}+1) - \Tr\left[\rho\frac{I + A_1}{2}\right]\right|\leq \epsilon$ with probability at most $1-\exp(-2\epsilon^2 N)$ from Hoeffding's inequality.
The same bound holds for every term in Eq. (\ref{eq:difference_rhoandmu}).
Noting that if $\left|\frac{1}{2N}\sum_{j=1}^N(b_{3j}+1) - \Tr\left[\rho\frac{I + A_1}{2}\right]\right|\leq \epsilon$ holds, $\left|\frac{1}{2N}\sum_{j=1}^N(1-b_{3j}) - \Tr\left[\rho\frac{I - A_1}{2}\right]\right|\leq \epsilon$ also holds, the probabilty that at least one of the differences in Eq. (\ref{eq:difference_rhoandmu}) is larger than $\epsilon$ is at most $4\exp(-2\epsilon^2 N)$, by union bound.
Therefore, with probability at least $1 - 4\exp(-2\epsilon^2 N)$, 
\begin{align}
    &\left\|\mathcal{S}(e^{i\theta A_1\otimes A_2})\kket{\rho} - \sum_{i=1}^6 c_i \kket{\bar{\mu_i}}\right\| \nonumber \\
    &\leq 
    \epsilon \left\| \sum_{i=3}^6 \sum_{\beta\in\{1,-1\}}c_i\Phi_{i, \beta}\kket{\rho}\right\| \nonumber \\
    &\leq 
    \epsilon \sum_{i=3}^6 \sum_{\beta\in\{1,-1\}}\left\| c_i\Phi_{i, \beta}\kket{\rho}\right\|,
\end{align}
holds for any norm $\|\cdot\|$.
The second inequality follows from the triangle inequality. 
Considering the trace norm $\|\cdot\|_1$, which gives $\|\Phi_{i,\beta} \kket{\rho}\|_1 = 1$, and taking $|c_i|\leq 1$ into account, we get
\begin{align}
    \left\|\mathcal{S}(e^{i\theta A_1\otimes A_2})\kket{\rho} - \kket{\bar{\mu_i}}\right\| &\leq 8\epsilon.
\end{align}

With this, we conclude that, given the desired error $1/\epsilon$ and a probability $1-\delta$ by which we wish to lower-bound the probability of getting the error larger than $\epsilon$, we can take $N= -\frac{32}{\epsilon^2}\ln (1-\delta)$.

\section{Time-like cut for identity channel}\label{sec:peng-time-like-cut}
The time-like cut approach proposed in Ref.~\cite{Peng2019} can be derived in the following manner.
Let us consider an identity channel $\mathcal{I}_a$ on the $a$-th qubit. it can be expanded as,
\begin{equation}
    \mathcal{I}_a = \sum_{i_a=0}^3 \sum_{j_a=0}^3 \kket{e_{i_a}}\bbra{e_{j_a}} \bbra{e_{i_a}}\mathcal{I}_a\kket{e_{j_a}}
\end{equation}
Since we assumed $\kket{e_i}$ are orthonormal to each other and $\mathcal{I}\kket{\rho} = \kket{\rho}$ for any $\rho$,
\begin{equation}
    \mathcal{I}_a = \sum_{i_a=0}^3 \kket{e_{i_a}}\bbra{e_{i_a}} 
\end{equation}
If we apply this to a $n$-qubit density matrix $\kket{\rho}=\sum_{j_1,\cdots, j_n} \rho_{\bm{j}} \kket{e_{j_1}e_{j_2}\cdots e_{j_n}}$ in this form, we see,
\begin{align}
    \mathcal{I}\kket{\rho} &= \sum_{i_a=0}^3 \kket{e_{i_a}}\bbraket{e_{i_a}|\rho}\\
    &= \sum_{i_a=0}^3 \Tr_a(e_{i_a}\rho) \otimes \kket{e_{i_a}}.
\end{align}
Choosing $\kket{e_{j_1}}$ to be Pauli matrices $\{I, X, Y, Z\}/\sqrt{2}$, we conclude,
\begin{align}
    \mathcal{I}\kket{\rho} &= \frac{1}{2}\sum_{A \in \{I,X,Y,Z\}} \Tr_a(A_a  \rho) \otimes A_a.
\end{align}
This equation implies that we can first measure expectation values of $X, Y, Z$ at the $a$-th qubit and then re-input each eigenstates.

\section{Analysis of the simple example given in Sec. \ref{sec:simulation}}\label{app:simpleexample}

\subsection{Cost of time-like cut}
First, we consider the ``time-like'' cut approach.
Let us name two-qubit on which the CZ gate acts $a$ and $b$, and let $\sigma_a$ and $\sigma_b$ be their 1-qubit reduced density matrices after the gate $U_1\otimes V_1$.

We assume,
\begin{itemize}
    \item One $n$-qubit device is available.
    \item Qubits that are measured in the basis of an observable $O_i$ can be reused to prepare the input state $\rho_j$.
\end{itemize}
We take the following naive approach to estimate $\expect{O_f}$.
First, divide the allowed number of circuit runs $N$ into $N/2$ to run the divided circuit for equal times.
$N/2$ runs are further divided into $N/128$ runs to run the circuit with $O_i$ and $\rho_j$ for $i, j \in \{1,2,\cdots,8\}^2$ \footnote{Note that this approach is almost equivalent for a large $N$ to the approach proposed in Ref. \cite{Peng2019}, which proceeds by drawing the pair $i,j$ from uniform distribution.}.
With each $N/128$ runs with a pair $(O_i, \rho_j)$, we estimate the value of the tensor network below.
\includegraphics[width=\linewidth]{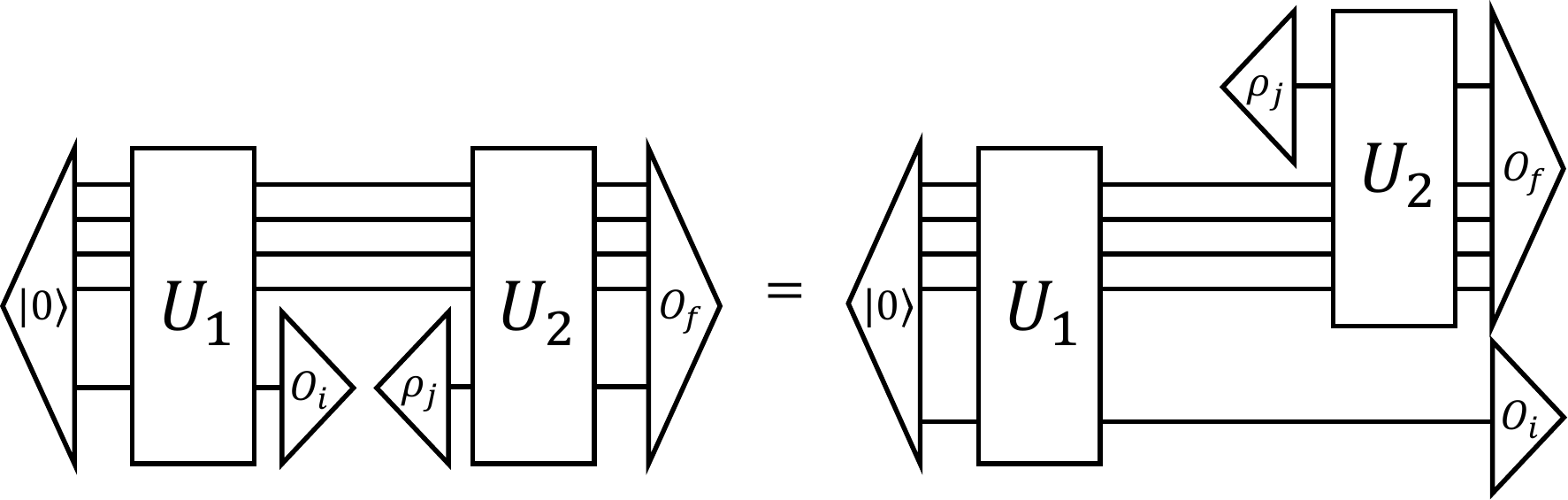}
Since we assumed $O_f$ is a tensor product of Pauli $Z$'s and $O_i$ is drawn from ${I,X,Y,Z}$, from each run we obtain a measurement result $o_{ij, r} = \pm 1$, where $r$ is the index to distinguish the runs.
Using $o_{ij,r}$, we estimate the above tensor network by $\tilde{o}_{ij} = \frac{1}{N/128}\sum_{r=1}^{N/128}o_{ij,r}$.
Since $o_{ij, r}$ is a random variable which takes $\{+1, -1\}$, this estimator $\tilde{o}_{ij}$ approximately follows a normal distribution with an expectation $\mathbb{E}[o_{ij}]$ and a variance $\frac{1}{N/128}(1-\mathbb{E}[o_{ij}]^2)\leq \frac{1}{N/128}$, for sufficiently large $N$.
Therefore, with $N/128$ runs of a quantum circuit, we can estimate the value of the above tensor network with the variance $\frac{128}{N}$ at most. 
We can obtain the same result for the other cluster.

For each $i,j,k,l$, the tensor in the sum of Fig.~\ref{fig:simpleclustering} is estimated by the product of the above estimators because the measurement result is independent on each cluster.
The variance of each tensor network can be evaluated because they are a product of two random variables approximately drawn from normal distribution with variance at most $\frac{128}{N}$, and it is at most $\frac{64}{N}$.
We have that $c_{ijkl}\in \{\pm 1/2^4\}$.
This reduces the variance of each term in the summation to $1/(2N)$.
However, when we take the sum since each term can be approximated by a normal distribution with the variance at most $1/(2N)$ and we take summation of $8^4 = 4096$ terms, the result has the variance at most $2048/N$.

\subsection{Cost of space-like cut}
For the space-like cut, we divide the allowed number of circuit runs, $N$, to $N/6$ \footnote{This is also a naive approach and there is a possibility to optimize the way of division, as mentioned in the main text.}.
First, we run eight circuits that do not involve the measurement in the middle and obtain estimators for these four tensor network.
Each of the estimators has the variance at most $6/N$.
Let us now move on to the circuits with the measurement.
For arbitrary density matrix $\rho'$, the $Z$ measurement produces the density matrix $\left(\frac{I+\alpha}{2}\right)\rho'\left(\frac{I+\alpha}{2}\right)/p_\alpha$ with probability $p_\alpha$.
Therefore, to obtain the above decomposition, we need to know the normalization factor $p_\alpha$.
With $N/6$ circuit runs, $p_\alpha$ is estimated to the variance $p_\alpha(1-p_\alpha)/(N/6)$ which is at most $6/(4N)=3/(2N)$.
Note that when $N$ is large, the distribution of the estimator $\tilde{p}_\alpha$ can be thought of as a normal distribution.
Conditioned on $\alpha$, we construct an estimator of $O_f$.
Since $\alpha$ is obtaned with probability $p_\alpha$, for each $\alpha$ we have $p_\alpha N/6$ samples to estimate $O_f$.
Therefore, for each $\alpha$, the estimator of $O_f$ has the variance of $6/p_\alpha N$.
Since the estimator of the tensor network is the product of the estimators of conditioned $O_f$ and $p_\alpha$, its variace is $\frac{6}{N}\frac{1}{p_\alpha + 1/[p_\alpha(1-p_\alpha)]} \leq \frac{3}{2N}$.
Each pair of the tensor network in Fig. \ref{fig:cz_gate_cut} is multiplied together, and if we perform this with the estimators obtained above, the variance of each term is at most $\frac{3}{N}$.
We further multiply each term with $\pm 1/2$, then the variance is reduced to $\frac{3}{4N}$.
Finally, the summation of 10 such term leads to the variance of $\frac{15}{2N}$.

\section{Proof of Theorem \ref{thm:multiple_twoqgate_cut}} \label{appsec:proof_of_multiple_twoqgate_cut}
We follow the approach taken in Ref. \cite{Peng2019}.
The task here is to perform the decomposition of $m$-qubit circuit like the one shown in Fig. \ref{fig:combination} so that the original quantum circuit can be approximated with an $n$-qubit quantum computer, where the input state $\rho$ is initialized in $\ket{0}\bra{0}^{\otimes m}$ and $O_f$ is an output (diagonal) observable calculated from some output function $f:\{0,1\}^m\to [-1,1]$.
We want to estimate $\mathbb{E}_y[f(y)]$ for $n$-bit measurement outcomes $y$ to some accuracy $\epsilon >0$ with some high probability $1-\delta$.

Let the number of space-like cuts and time-like cuts performed in the decomposition be $M_s$ and $M_t$ respectively.
We assume the space-like cuts are performed only on CZ gates.
We redefine the probabilistic map $\Phi'_i$ that is used to decompose $\mathcal{S}(e^{i\theta A\otimes B})$ as,
\begin{align}
    \Phi_{1}' &= \mathcal{S}(I\otimes I), \nonumber \\
    \Phi_{2}' &= \mathcal{S}(A\otimes B), \nonumber \\
    \Phi_{3}' &= \mathcal{M}_{A} \otimes \mathcal{S}(e^{i\pi B/4}), \nonumber\\
    \Phi_{4}' &= \mathcal{M}_{A} \otimes \mathcal{S}(e^{-i\pi B/4}), \\
    \Phi_{5}' &= \mathcal{S}(e^{i\pi A/4}) \otimes \mathcal{M}_{B}, \nonumber \\
    \Phi_{6}' &= \mathcal{S}(e^{-i\pi A/4}) \otimes \mathcal{M}_{B}, \nonumber
\end{align}
where $\mathcal{M}_{A}$ and $\mathcal{M}_{B}$ and the projective measurement of $A$ and $B$.
Let $s_k\in \{1,\cdots,6\}$ be an index of the above probabilistic map $\Phi'_{s_k}$ applied to the $k$-th space-like cut $k\in\{1,\cdots,M_s\}$ and $t_l$ be an index of an observable-state pair $(O_{t_l}, \rho_{t_l})$ in Eq. (\ref{eq:Pengcut_pairs}) applied to the $l$-th time-like cut $l\in\{1,\cdots,M_t\}$.
The coefficients associated with a space-like cut (Eq. (\ref{eq:space-like-cut-coefs})) and a time-like cut (Eq. (\ref{eq:Pengcut_pairs})) are redefined as $c_{s_k}^{\mathrm{space}}$ and $c_{t_l}^{\mathrm{time}}$, respectively.
With one set of indices, $s=\{s_k\}_{k=1}^{M_s}\in\{1,\cdots,6\}^{M_s}$ and 
$t=\{t_k\}_{k=1}^{M_t}\in\{1,\cdots,8\}^{M_t}$, we can define a corresponding quantum circuit which is induced by replacing every cut two-qubit gate by $\Phi_{s_k}'$ and every cut qubit line by the measurement of $O_{t_l}$ and the preparation of $\rho_{t_l}$.

When we run this circuit on $n$-qubit quantum device, we get the measurement outcomes at each cut, which is a string of $\pm 1$ from $\Phi_{s_k}'$ and the measurement of $O_{t_l}$, and the ones at the output qubit which is a bitstring of length $n$.
Let such outcomes from the $k$-th space-like cut, the $l$-th time-like cut and the output qubits be $b_{s_k}^{\mathrm{space}}\in \{+1, -1\}$, $b_{t_l}^{\mathrm{time}}\in \{+1, -1\}$ and $y_{(s,t)}\in\{0,1\}^n$, respectively.
Since $s_k=1,2$ does not involve measurement, we define $b_1^{\mathrm{space}}=b_2^{\mathrm{space}}=1$.
With the definition above and the equality for performing the decomposition (Eqs. (\ref{eq:space-like-cut-decomposition-with-prob-map}) and (\ref{eq:Pengcut_pairs}), Figs. \ref{fig:two_qubit_gate_cut} and \ref{fig:identitygatecut}), notice that,
\begin{widetext}
\begin{equation}\label{eq:combination-of-space-time-cut}
    \mathbb{E}_y[f(y)] = \sum_{s\in\{1,\cdots,6\}^{M_s}} \sum_{t\in\{1,\cdots,8\}^{M_t}} \prod_{k=1}^{M_s} c_{s_k} \prod_{l=1}^{M_t} c_{t_l} \mathbb{E}_{(\{b_{s_k}^{\mathrm{space}}\}, \{b_{t_l}^{\mathrm{time}}\}, y_{(s,t)})}\left[\prod_{k=1}^{M_s} b_{s_k}^{\mathrm{space}} \prod_{l=1}^{M_t} b_{t_l}^{\mathrm{time}} f(y_{(s,t)})\right],
\end{equation}
where the expectation on the right hand side is defined over a distribution of $\{b_{s_k}^{\mathrm{space}}\}, \{b_{t_l}^{\mathrm{time}}\}$ and $y_{(s,t)}$ for a quantum circuit induced by a given set of indices $(s,t)$.

We can take a Monte-Carlo approach to estimate the sum of the right-hand side of Eq. (\ref{eq:combination-of-space-time-cut}).
If we sample $s$ and $t$ from a uniform distribution on $\{1,\cdots,6\}^{M_s}$ and $\{1,\cdots,8\}^{M_t}$ respectively, Eq. (\ref{eq:combination-of-space-time-cut}) can be rewritten as,
\begin{equation}\label{eq:combination-of-space-time-cut-randomized}
    \mathbb{E}_y[f(y)] = \mathbb{E}_{(s,t,\{b_{s_k}^{\mathrm{space}}\}, \{b_{t_l}^{\mathrm{time}}\}, y_{(s,t)})} \left[6^{M_s}8^{M_t}\prod_{k=1}^{M_s} c_{s_k} b_{s_k}^{\mathrm{space}} \prod_{l=1}^{M_t} c_{t_l} b_{t_l}^{\mathrm{time}}  f(y_{(s,t)})\right],
\end{equation}
\end{widetext}
Let us define a random variable
\begin{equation}
    X_{(s,t)} = 6^{M_s}8^{M_t}\prod_{k=1}^{M_s} c_{s_k} b_{s_k}^{\mathrm{space}} \prod_{l=1}^{M_t} c_{t_l} b_{t_l}^{\mathrm{time}}  f(y_{(s,t)}).
\end{equation}
Let ${(s^{(i)}, t^{(i)})}_{i=1}^N$ be $N$ randomly sampled $(s,t)$ pair.
Then, $\mathbb{E}_y[f(y)]$ can be estimated by $\frac{1}{N}\sum_{i=1}^N X_{(s^{(i)}, t^{(i)})}$.
We will use the Hoeffding's inequality to bound the error of this Monte-Carlo approach.
The magnitude of $X_{(s,t)}$ is bounded by,
\begin{equation}\label{eq:magnitude-bound}
    \left| 6^{M_s}8^{M_t}\prod_{k=1}^{M_s} c_{s_k} b_{s_k}^{\mathrm{space}} \prod_{l=1}^{M_t} c_{t_l} \prod_{k=1}^{M_s} b_{t_l}^{\mathrm{time}}  f(y_{(s,t)})\right| \leq 3^{M_s}4^{M_t},
\end{equation}
because $|c_{t_l}|=1/2$, $|f(y_{(s,t)})|\leq 1$, $|b_{s_k}^{\mathrm{space}}|=1$, $|b_{t_l}^{\mathrm{time}}|=1$, and $|c_{s_k}|=1/2$ which follows from the assumption that the space-like cuts are performed only on CZ gates.
With the above bound of the magnitude, the Hoeffding's inequality guarantees that,
\begin{align}
    &\mathrm{Pr}\left[\left|\frac{1}{N}\sum_{i=1}^N X_{(s^{(i)}, t^{(i)})} - \mathbb{E}_y[f(y)]\right|\leq \epsilon\right] \\
    &\geq 1 - 2\exp\left(-\frac{N\epsilon^2}{2\cdot 9^{M_s}\cdot 16^{M_t}}\right).
\end{align}
Therefore, for given $\epsilon$ and the probability $1-\delta$ to which we want to bound the probability of getting an error larger than $\epsilon$, we take $N=\frac{2\cdot 9^{M_s}\cdot 16^{M_t}}{\epsilon^2}\ln\left(\frac{1}{2\delta}\right)$.

\end{document}